# Fullerenes and endohedrals as "big atoms"

## M. Ya. Amusia[1,2]

[1]Racah Institute of Physics, the Hebrew University, Jerusalem 91904, Israel
[2]Ioffe Physical-Technical Institute, St.-Petersburg 194021, Russia

**Abstract**

We present the main features of the electronic structure of the heavy atoms that is best of all seen in photoionization. We acknowledge how important was and still is investigation of the interaction between low- and high frequency lasers with big intensity. We discuss the fullerenes and endohedrals as big atoms concentrating upon their most prominent features revealed in photoionization. Namely, we discuss reflection of photoelectron wave by the static potential that mimics the fullerenes electron shell and modification of the incoming photon beam under the action of the polarizable fullerenes shell. Both effects are clearly reflected in the photoionization cross-section.

We discuss the possible features of interaction between laser field of both low and high frequency and high intensity upon fullerenes and endohedrals. We envisage prominent effects of multi-electron ionization and photon emission, including high-energy photons. We emphasize the important role that electron exchange can play in these processes.

*Endohedrals ionization and radiation*

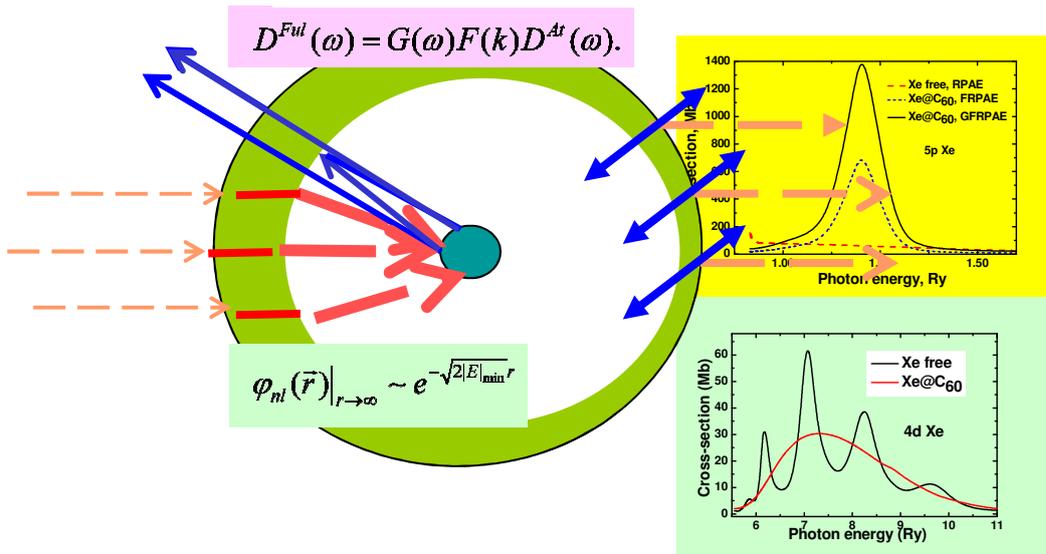

## 1. Introduction

Electronic structure of atoms is determined by shell effects and Coulomb interaction between electrons. An essential part of this interaction is included into the so-called self-consistent Hartree-Fock (HF) field and taken into account by using as one-electron HF wave-functions. However the remainder part of it, so-called direct or residual interaction leads to prominent effects particularly clear seen in photoionization cross-section with its Giant and Interference resonances. To describe atomic photoionization random phase approximation with exchange (RPAE) was introduced. With its help total and differential in angle photoionization cross-sections were described with accuracy compatible to that achieved in experiment. Atoms due to the well-known nature of its interparticle interaction proved to be excellent objects to check the accuracy of suggested theoretical approaches, demonstrating for example the abilities as well as limits of RPAE and the necessity in describing inner and in some cases even intermediate shells to go beyond RPAE frame by developing its generalizations GRPAE (see [1] and [2]).

Starting from pioneering experiments in 1975-77 on two-electron ionization of strontium and barium atoms in the intense by that time laser field ($10^{14}$ Watts/cm$^2$) [3], it became clear that some interesting features of isolated atoms can strongly affect the multi-electron photoionization process. Experiments demonstrated also that formation of ions in high intensity laser field is not limited by two-electron photoionization. On the contrary, it appeared that creation of multiply charged ions is a very probable process [4]. It was found that the achieved degree of ionization demonstrates the efficiency of simultaneous absorption of not several but hundreds of photons. Ordinary mechanisms like perturbation theory in laser field are absolutely inadequate in explaining the observed data.

Approximately at the same time, photons were detected [5] that could be emitted only after ionization of an intermediate or inner atomic subshell that requires ten and more photons with $\omega \cong 6.42 eV$ [1] of the laser used in [5].

The situation at that time became an object of intensive discussions in my group. We tried to apply ideas of strong intershell interaction, known to us from the atomic photoeffect studies, as a mechanism of multiple ionizations[2]. This direction of thought failed to give proper explanation, but contributed to the efforts that resulted in invention of the so-called "atomic antenna" mechanism [6]. The main point was the understanding that even a single ionized electron in the laser field with frequency $\omega$ acquires energy $E \sim \Upsilon / \omega^2$, where $\Upsilon$ is the intensity of the laser beam. Oscillating in this field, the electron will hit back the parent atom knocking out a second electron, if its energy exceeds the ionization potential of the remained ion $I_i$. For high enough $E$ even direct ionization of several electrons by the oscillating one is possible. Such a process could be repeated times and again, thus leading to indirect multi-electron ionization. Note that the energy of oscillating electron $E$ was called *pondermotive* in [7], where the mechanism [6] was rediscovered.

I was very much impressed by the simplicity and elegance of the picture and the efficiency with which the antenna mechanism can generate high energy electrons. It was immediately clear to me that in the same way radiation can be generated. Indeed, the oscillating electron while colliding with the parental ion can emit light in a process that can be called Internal Bremsstrahlung. This bremsstrahlung can be strongly affected by polarization of the target ion in the process of collision, leading to polarization or "atomic" internal Bremsstrahlung, just like in ordinary electron-atom collisions [8]. Obviously, intensity of emission of secondary electrons and radiation can be considerably enhanced due to presence of autoionization resonances in the oscillating electron - ion scattering cross-section (see model in [9]).

---

[1] Atomic system of units is used in this paper: electron charge *e*, its mass *m* and Plank constant $\hbar$ are equal to 1, $e = m = \hbar = 1$

[2] I am grateful to N. B. Delone, whose enthusiasm inspired attention to this field of a number of people, including myself.

I vividly remember, that it became also immediately clear that for efficiency of non-linear processes the key-point is the frequency of the laser – the smaller it is the higher energy could be generated. As a possible candidate microwave radiation was considered. But it appeared that the respective amplitude of electron oscillation $a_e$ is too big, since to achieve sufficient for atomic ionization $E$ the radius $a_e \sim \sqrt{\Upsilon}/\omega^2$ became much bigger than the interparticle distance at the best available vacuum.

An interesting idea appear that not one, but a number of ionized electrons $N_C$ liberated one after another off the atom due to re-scattering process can collectively oscillate and collide with the parental ion increasing the amplitude of electron or photon emission by a factor that is proportional to $N_C^2$.

The ideas of [6] and some other mentioned above were presented by me in 1989 at a round table on laser physics held during an ICPEAC meeting in New York and were accepted without a trace of interest. The common view was summarized by the chairman, a very well-known laser theorist: "This is so simple that if it would be correct, everybody would know it. So, this is incorrect".

Irrespective to this opinion, the described mechanism became a generally accepted fundament for understanding of the low frequency high intensity lasers interaction with atoms.

It seemed at least to me that when going to high enough frequencies the situation will become simpler and perturbation theory in the laser field will be sufficient. Now sources of laser radiation, so-called free electron lasers (FEL), with the photon energy of about 90.5 eV [10] became available. For them the pondermotive energy is very small as compared to atomic ionization potential. However, in the studies of its interaction with Xe were produced highly charged ions, up to $Xe^{+19}$ [10]. It became evident that the field is not exhausted and some other ideas may be of importance to describe FEL interaction with atoms.

Similar to atoms microscopic, in fact, Nano-scale size objects in the field of laser-target interaction became available relatively recently. I have in mind not only noble atoms' clusters that were already used at the very beginning of the nineties [11] (see e.g. also [12]), but metallic clusters and particularly fullerenes and endohedrals. It seems that very attractive is the fullerene $C_{60}$, and corresponding noble gas endohedrals $NG@C_{60}$. They have very powerful resonances in photoionization cross-section that we will discuss later in this paper and a sort of a nucleus the role of which in $NG@C_{60}$ is played by the noble gas atom. These "quasi-nuclear" reactions we will also discuss in this paper.

At the end of eighties, I start to question the generally accepted method of estimation of the inner electron direct ionization probability. Namely, I came to the conclusion that the asymptotic of inner electron wave functions was estimated incorrectly. From direct calculations we knew and have analytic confirmations (see also [13, 14]) that when exchange between outer and inner electrons (it is in Hartree-Fock approximation) is taken into account the inner electron wave function asymptotic acquire admixture of the outer electron.

On the ground of this asymptotic behavior it was demonstrated that inner shell ionization in a strong electric field is by many orders of magnitude bigger than estimated in the frame of an ordinary one-electron approximation [15]. The reaction of colleagues, first of all late Prof. U. Fano, to this idea was lukewarm. This is why a paper on this subject was not written for many years, although I did not found any defects in own argumentation. Perhaps, I would never present it as a paper if not recently run across the e-prints [16, 17]]. This was the big straw that pushed me at last in the right, as I believe, direction to publish these results [18]. This approach could be of importance and I will discuss the effect of electron exchange upon the ionization of atoms, fullerenes and endohedrals in a laser field. I will present briefly some arguments against the objections against the possible role of "exchange mechanism" that was presented in [19][3].

In general, the main point of this paper is to show that fullerenes and endohedral, being similar to very big atoms, are very promising objects for

## 2. One-electron and collective effects in atomic photoionization

As one-electron, most accurate in describing an isolated atom (ion) with nuclear charge $Z$ and total number of electron $N$ is the well-known Hartree-Fock approximation, in the frame of which the wave functions of each of electrons is determined by solving the following system of equations (see, e.g. [1])

$$\left[-\frac{\Delta}{2}-\frac{Z}{r}+\sum_{k=1}^{N}\int \rho_k(x')\frac{dx'}{|\vec{r}'-\vec{r}|}\right]\varphi_j(x) - \sum_{k=1}^{N}\int \varphi_k^*(x')\frac{dx'}{|\vec{r}'-\vec{r}|}\varphi_j(x')\varphi_k(x) \equiv$$
$$\equiv \hat{H}_H \varphi_j(x) - \sum_{k=1}^{N}\int \varphi_k^*(x')\frac{dx'}{|\vec{r}'-\vec{r}|}\varphi_j(x')\varphi_k(x) \equiv \hat{H}_{HF}(x)\varphi_j(x) = E_j \varphi_j(x)$$
(1)

where subscript $H$ ($HF$) stands for Hartree (Hartree-Fock), $x = \vec{r},\vec{s}$ is the coordinate and spin projection, $\rho_k(x) \equiv |\varphi_k(x)|^2$ is the one-electron $k$ state density. The total electron $\rho(x)$ density is given by $\rho(x) = \sum_{k=1}^{N}\rho_k(x)$ It is seen from (1) that at large distances the effective potential behaves as $(-Z+N-1)/r$.

The contribution of the second term in the integrand cannot be presented as an action of some local potential $W(r)$ upon $\varphi_j(x)$. On the contrary, the action described by this term is non-local, connecting points $x$ and $x'$, over which the integration is performed.

Closed shell atoms are spherically-symmetric objects. For them it is precise that $k \equiv n(\varepsilon), l, m, s$, where $n(\varepsilon)$ is the electron principal quantum number (continuous energy), $l$ is the angular momentum, $m$ is its projection, and $s$ is the spin projection.

To describe photoionization, including that of inner and intermediate subshells of even heavy atoms, the so-called dipole approximation is accurate enough. Then in HF the photoionization amplitude $d_{if}$ that describes atomic electron transition from the initial $i$ into final $f$ states is given by the following expression

$$d_{i \to f} \equiv \langle f|\hat{d}|i\rangle = \int \varphi_f^*(x)d(x)\varphi_i(x)dx$$
(2)

where $\hat{d} = d(x)$ is the operator that describes dipole part of photon-electron interaction.

To take into account inter-electron residual interaction the so-called random-phase approximation (RPAE) is widely used. It preserves some features of HF but takes into account the response for time-dependence of individual electron states. In RPAE the equation that determines the photoionization amplitude $D_{nl \to \varepsilon l'} = \langle \varepsilon l'|D(\omega)|nl\rangle$ for a closed subshell atom is presented as [1]

---
[3] I am grateful to Dr. V. Averbukh, who in September 2011 attracted my attention to [19] that I was not aware of.

$$\langle \varepsilon l' | D(\omega) | nl \rangle = \langle \varepsilon l' | \hat{d} | nl \rangle +$$
$$+ \left( \sum_{\varepsilon''l'' \leq F, \varepsilon'''l''' > F} - \sum_{\varepsilon'''l''' < F, \varepsilon''l'' \leq F} \right) \frac{\langle \varepsilon'''l''' | D(\omega) | \varepsilon''l'' \rangle \langle \varepsilon''l'', \varepsilon l' | U | \varepsilon'''l''', nl \rangle}{\omega - \varepsilon_{\varepsilon'''l'''} + \varepsilon_{\varepsilon''l''} + i\eta(1 - 2n_{\varepsilon''l''})} \quad (3)$$

Note that in this case an electron state is sufficient to characterize by two instead of four quantum numbers, namely $k \equiv n(\varepsilon), l$ and $l' = l \pm 1$. In (3) $\leq F(>F)$ denotes summation over occupied (vacant) atomic levels in the target atom. Summation over vacant levels includes also integration over continuous spectrum, $n_{\varepsilon l}$ is the Fermi step function that is equal to 1 for $nl \leq F$ and 0 for $nl > F$; the Coulomb inter-electron interaction matrix element is defined as $\langle \varepsilon''l'', \varepsilon l' | U | \varepsilon'''l''', nl \rangle = \langle \varepsilon''l'', \varepsilon l' | r_< / r_>^2 | \varepsilon'''l''', nl \rangle - \langle \varepsilon''l'', \varepsilon l' | r_< / r_>^2 | nl, \varepsilon'''l''' \rangle$ and $\eta \to +0$. In the latter formula notation of smaller (bigger) radiuses $r_<(r_>)$ of interacting electron coordinates comes from the well-known expansion of the Coulomb inter-electron interaction. The necessary details about solving (3) one can find in [20].

To analyze the equation (3), it is convenient to present it in a symbolic, operator form [2]:

$$\hat{D}(\omega) = \hat{d} + \hat{D}(\omega)\hat{\chi}(\omega)\hat{U}; \quad \hat{\chi}(\omega) \equiv 1/(\omega - \hat{H}_{exc}) \quad . \quad (4)$$

Here $\hat{H}_{exc}$ is the electron excitation or electron-vacancy creation Hamiltonian in HF approximation $\hat{\chi}(\omega)$ is called electron-vacancy propagator. The equation (4) can be easily solved symbolically, leading to

$$\hat{D}(\omega) = \hat{d} / [1 - \hat{\chi}(\omega)\hat{U}] \quad (5)$$

The amplitude $\hat{D}(\omega)$ is enhanced as compared to $\hat{d}$ when the denominator in (5) becomes small. If for some $\Omega > I$, where $I$ is the atomic ionization potential, the condition $1 - \hat{\chi}(\Omega)\hat{U} = 0$ is fulfilled, the photoionization cross-section has a Giant resonance of multi-electron nature. The best example of it is the powerful maximum in $4d^{10}$ photoionization cross-section of Xe with the total oscillator strength of about 10 (see Fig.1).

RPAE helped to predict and discover another sort of multi-electron resonances that we call now *interference* resonances. To describe them, let us consider a situation, in which the direct HF amplitude $d_s$ is small, while there are other electrons with big photoionization amplitude $D_b$, $\hat{D}_b(\omega) \gg d_s$. Then from (4) one has

$$\hat{D}_s(\omega) \approx d_s + \hat{D}_b(\omega)\hat{\chi}(\omega)\hat{U}_{bs} \approx \hat{D}_b(\omega)\hat{\chi}(\omega)\hat{U}_{bs} \gg d_s, \quad (6)$$

if the *inter-transition* interaction $U_{bs}$ is not too small. The enhancement of the photoionization amplitude described by (6) manifests itself as a resonance in the partial cross section of *s* electrons photoionization. Very often the term $\hat{D}_b(\omega)\hat{\chi}(\omega)\hat{U}_{bs}$ and $d$ have opposite signs, so the total amplitude acquires along with an extra maximum two minima, thus forming in the partial cross section a rather complicated structure that was named *interference* or *correlation* resonance (see [2]).

Several other resonances were predicted in the frame of RPAE and later found in experiment [2].

## 3. Photoionization of Fullerenes and Endohedrals by single photons

Of the all different fullerenes, we will concentrate here on $C_{60}$. Amazing is its perfect, close to spherical symmetry, shape and its thin, as compared to its radius $R_C$ shell. The remarkable feature is that $R_C \approx 6.4$ is considerably bigger than the atomic radius $R_A \simeq 1$. It is essential that the thickness of the fullerenes shell $\Delta_C \simeq 1.5 \div 2$ is much smaller than $R_C$. In relatively crude but reasonable approximation $C_{60}$ can be considered as a spherical shell that include 240 strongly collectivized electrons and 60 ions $C^{+4}$ that can be mimicked by a positively charged spherical shell. There exist other than $C_{60}$ fullerenes but their form is not spherical and they present much more complex objects for at least theoretical photoionization studies.

The photoionization cross-section in $C_{60}$ is dominated by a powerful maximum, so-called Giant resonance [21]. It maximum is located at about 2 Ry and total oscillator strength is close to 240 that is more than twenty times stronger than in $4d^{10}$ Xe. Direct application of equations (1) and (3) to $C_{60}$ is very complex, but the very fact that Giant resonance in $C_{60}$ was predicted [22] on the bases of solving a schematic model similar to RPAE leaves no doubt that (3) is suitable to describe $C_{60}$. Fullerene presents a sort of an atom, with developed shell structure and impressive collective effects in photoionization, however without a concentrated nucleus. Note that $C_{60}$ has a quite high polarizability that transforms, in fact, focuses the electromagnetic field inside $C_{60}$ as compared to the external one.

In this sense endohedrals that present a fullerene, stuffed by an atom, is essentially different. It appeared that inside $C_{60}$ and bigger fullerenes almost any atom of the periodic table or even small molecules can be located. Note, that in bigger fullerenes, like $C_{240}$ one can stuff $C_{60}$ itself. A fullerene with $N_C$ carbon atoms and an atom A inside is denoted as $A@C_{N_C}$. Detailed description of different aspects of photoionization of $A@C_{N_C}$ can be found in [23].

Of course, there are different features in big atoms - endohedrals that are absolutely inessential in their photoionization. For instance, the shell structure of $A@C_{N_C}$ is weakly affected by the atom A. The "nucleus" A can be either a little bit stretched or compressed just as the shell $C_{N_C}$ by atom's A action. These effects are even much weaker in real atoms where the nucleus is smaller than the atomic shell by five orders of magnitude. But due to proximity of A and the shell $C_{N_C}$, contrary to the case of ordinary atoms, processes in A and $C_{N_C}$ in $A@C_{N_C}$ are closely interconnected.

We will concentrate on endohedral $A@C_{60}$ where A is a noble gas atom, from He to Xe, and investigate the effect of the unperturbed by A the $C_{60}$ shell. The most important are the following effects of $C_{60}$ upon photoionization cross-section of atom A located inside $C_{60}$. The first is the elastic and inelastic scattering of photoelectrons from A by $C_{60}$ while the second is the modification of the electromagnetic field acting upon A due to action of the fullerenes shell (see [24] and references therein).

In describing these two effects we rely upon RPAE, adopting it to $A@C_{60}$ by introducing two factors: photoelectron scattering and photon beam modification. The real inequality $R_C > R_A \approx \Delta_C$ is substituted by a model, simplifying inequality $R_C \gg R_A \gg \Delta_C$. In this approach the photoionization amplitude of an electron from $A@C_{60}$ $nl$ subshell $D_{nl \to \varepsilon l \pm 1}^{AC}(\omega)$ in RPAE with account of reflection of photoelectrons by the $C_{60}$ shell and polarization of the latter under the action of the incoming photon beam can be presented as the following product [24, 25]

$$D^{AC}_{nl \to \varepsilon l \pm 1}(\omega) = G(\omega) F_{l \pm 1}(k) D^{F}_{nl \to \varepsilon l \pm 1}(\omega). \tag{7}$$

Here the polarization factor $G(\omega)$ takes into account the modification of the incoming photon beam by the fullerene $C_{N_C}$, $F_{l \pm 1}(k)$ describes the reflection factor that represent the effect of the fullerenes shell $C_{N_C}$ upon the outgoing photoelectron with the angular momentum $l \pm 1$ and linear momentum $k$, energy $\varepsilon = k^2/2$, connected to the photon frequency $\omega$ by the relation $\varepsilon = \omega - I_{nl}$, where $I_{nl}$ is the $nl$ subshell ionization potential. In (7) $D^{F}_{nl,\varepsilon l \pm 1}(\omega)$ is the atomic photoionization amplitude, in which the virtual states are modified due to action of the static potential of the fullerenes shell upon the virtually excited atomic states.

To obtain the reflection factor $F_{l \pm 1}(k)$, we substitute the fullerenes shell action by a static zero-thickness potential [26]

$$W(r) = -W_0 \delta(r - R). \tag{8}$$

The parameter $W_0$ is defined from the condition that the binding energy of extra electron in negative ion $C_{60}^{-}$ is equal to the experimentally observed value. The factor $F_{l'}(k)$ is determined by the expression [26, 25]:

$$F_l(k) = \cos \Delta \delta_l(k) \left[ 1 - \tan \Delta \delta_l(k) \frac{v_{kl}(R)}{u_{kl}(R)} \right], \tag{9}$$

where $\Delta \delta_l(k)$ is the addition to the photoelectron elastic scattering phase of the partial wave $l$ due to action of the potential (8), $u_{kl}(r)$ is the regular and $v_{kl}(r)$ irregular at point $r \to 0$ radial parts of atomic Hartree-Fock one-electron wave functions. The following relation expresses the additional phase shift $\Delta \delta_l(k)$:

$$\tan \Delta \delta_l(k) = \frac{u_{kl}^2(R_C)}{u_{kl}(R_C) v_{kl}(R_C) - k/2W_0}. \tag{10}$$

The factor $F_{l'}(k)$ as a function of $k$ oscillates due to interference between the direct photoelectron wave and its reflections from the fullerenes shell. This factor redistributes the resulting cross section as compared to that of the isolated atom but cannot change its value integrated over essential $\omega$ region.

We obtain $D^{F}_{nl,\varepsilon l \pm 1}(\omega)$ in the frame of the RPAE. When the fullerene shell is presented by the potential (8), the following equation is valid instead of (3)[25]:

$$\langle \varepsilon l' | D^F(\omega) | nl \rangle = \langle \varepsilon l' | d | nl \rangle +$$

$$+ \sum_{\varepsilon'l',\varepsilon''l''} \frac{\langle \varepsilon'''l''' | D^F(\omega) | \varepsilon''l'' \rangle \left[ F^2_{\varepsilon''l''} n_{\varepsilon'l''}(1 - n_{\varepsilon''l''}) - F^2_{\varepsilon'l'} n_{\varepsilon''l''}(1 - n_{\varepsilon'l'}) \right]}{\omega - \varepsilon_{\varepsilon''l''} + \varepsilon_{\varepsilon'l'} + i\eta(1 - 2n_{\varepsilon''l''})} \times . \quad (11)$$

$$\times \langle \varepsilon''l'', \varepsilon l' | U | \varepsilon'''l''', nl \rangle.$$

Here $\langle \varepsilon l' | D^F(\omega) | nl \rangle \equiv D^F_{nl,\varepsilon l \pm 1}(\omega)$. The approximation described by (11) is called FRPAE.

As is seen from (9), as a function of $k$ $F_{l\pm 1}(k)$ oscillates leading in photoionization cross-section to so-called *confinement resonances* (see, e.g. [27]). An example of them for 4d electrons in $Xe@C_{60}$ is presented in (Fig.1), along with the atomic 4d Giant resonance.

Now let us concentrate on the effect of polarization of fullerenes electron shell. Taking into account that the atom's A radius $R_A$ is considerably smaller than the fullerenes radius $R_C$, a rather simple expression can be obtained for $G(\omega)$

$$G(\omega) = 1 - \frac{\alpha_C(\omega)}{R_C^3}. \quad (12)$$

Here $\alpha_C(\omega)$ is the dipole polarizability of the fullerenes shell. Formula (12) was derived in [28] under simplifying assumption that $R_A / R_C \ll 1$. While $\alpha_C(\omega)$ is difficult to calculate ab-initio, it can be easily expressed via experimentally quite well known photoionization cross-section $\sigma_C(\omega)$ of the $C_{60}$ (see [21] and references therein):

$$\operatorname{Re}\alpha_C(\omega) = \frac{c}{2\pi^2} \int_{I_F}^{\infty} \frac{\sigma_C(\omega')d\omega'}{\omega'^2 - \omega^2}; \quad \operatorname{Im}\alpha_C(\omega) = c\sigma_C(\omega)/4\pi\omega. \quad (13)$$

Here $I_C$ is the fullerene ionization potential and $c$ is the speed of light.

Since the cross-section $\sigma_C(\omega)$ is absolutely dominated by the fullerenes Giant resonance that have a maximum at about 2Ry, $G(\omega)$ starts to decreases rapidly at $\omega > 2Ry$ reaching its asymptotic value equal to 1 at about 5Ry. This factor, connecting the atomic and fullerenes photoionization cross-section, is able to alter considerably the endohedral cross-section as compared to pure atomic one.

The polarization factor $G(\omega)$ has its own maxima that lead to *polarization resonances*. Combinations of confinement and polarization resonances lead to *Giant endohedral resonances* typical for outer subshells of atom A. This is exemplified in Fig. 2 by the 5p electrons of Xe in $Xe@C_{60}$ [29] where GRPAE denotes cross-section calculated using amplitude (7). We see formation of resonances that were called Giant confinement resonances tht show up in the outer shell close to its threshold.

The credibility to the given in this section rather crude approach to RPAE equations was added when the prediction of strong variation of 4d Giant resonance in $Xe@C_{60}$ as compared to pure Xe made in [30] was confirmed semi-quantitatively in [31], as is seen in Fig.1 [25]. It is of interest to note that confinement resonances in an inner or intermediate subshell can affect the outer shell even above the inner threshold. This is illustrated by the example of 5p cross section above 4d threshold in $Xe@C_{60}$ that is presented in Fig.3.

## 4. Fullerenes and Endohedrals in strong laser fields

As it was mentioned in the Introduction, recent experiments demonstrated the strong effect of atomic giant resonance at 8 Ry in Xe on formation of multiply charged ions, up to $Xe^{+19}$, in a field, for which the pondermotive energy is small [10]. Fullerene $C_{60}$ has a much more powerful resonance at 2 Ry and it is reasonable to expect much more impressive manifestation of it than in Xe.

Fullerenes are natural and promising objects for studies of their interaction with low frequency and high intensity lasers, for which the pondermotive energy is much bigger than the ionization potential $I_{C_N}$ i.e. $(\Upsilon/\omega^2 I_{C_N}) \gg 1$. Since the ionization potential in $C_N$, contrary to atoms, weakly depends upon the degree of ionization, it requires much less photons to liberate the same number of electrons than from an isolated atom. In addition the fullerene size is much bigger than that of an atom, so collisions with the target while initially ionized electron bounces around became much more probable than even for heavy atoms. And the total number of the electrons that are able to participate actively in laser-fullerene interaction process is much bigger than in the laser-atom interaction. As a result, laser-fullerene interaction could produce not only many high energy electrons but generate electromagnetic radiation of high energy photons with intensity much bigger than in laser-atom interaction.

Note that emission of radiation due to internal bremsstrahlung is enhanced not only due to autoionization resonances whose oscillator strength is as a rule considerably less than one unit but by Giant resonances with their combined oscillator strength of about ten units in atoms and several hundred units and more in fullerenes and endohedrals.

One should have in mind that fullerenes have high particle density in the surface dimension while is empty space inside. It means that the electrons that oscillate in the laser field can have a pondermotive radius $a_e \sim \sqrt{\Upsilon}/\omega^2$ can be $a_e \simeq R_{C_N}$ i.e. much bigger that the distance between two neighbor carbon atoms. It permits to acquire pondermotive energy that is much bigger than in the case of interaction with solid state objects.

Relatively long ago it was suggested that electron transitions in atomic shells could stimulate nuclear excitations (see e.g. [32]). The effect, however, was so small that it took years of research and no decisive answer was found yet. The main problem here is the fact that atomic transition of reasonable probability are dipole and their energy is very small in the nuclear scale. To have sufficiently strong interaction between atomic and nuclear transitions, their energies has to be close to each other. However, low energy nuclear transitions are usually non-dipole. But interaction between transitions with different angular momentum is very small.

The situations dramatically changes when the atom is excited by a strong laser field. If the inequality $(\Upsilon/\omega^2 I_{C_N}) \gg 1$ is fulfilled, laser ionized electrons can acquire MeV energies already at $\Upsilon \geq 10^{18} Watt/cm^2$, thus easily generating reaction in the nucleus of the atom. The situation becomes much simpler when the oscillating electrons collide with the "quasi-nucleus" in endohedrals. When an electron generated in interaction of a laser beam with intensity $\Upsilon \geq 10^{16} Watt/cm^2$ from $C_{60}$ in $Xe@C_{60}$ collides with Xe in the process of re-scattering, vacancies in 1s shell of Xe can be created, leading to photons with $\omega \simeq 30$ keV ! Needless to say that each 1s vacancy creation is followed by an avalanche of decays and production of very many electrons.

Thus, it is seen that fullerenes and endohedrals are quite interesting objects to study their interaction with high intensity lasers.

## 4. Exchange mechanism of ionization by a static field

Here we with comment on how essentially the exchange term in (3) modifies the long distance behavior of the inner electron wave function and thus its probability of under barrier escape or ionization [18]. Indeed, the exchange with the outer electron leads to big increase of the asymptotic value that determines the probability of ionization under the action of static field.

It is well known that in an attractive spherically – symmetric potential the asymptotic of the wave function $\varphi_{nl}(\vec{r})$ is determined by the binding energy $E_{nl}$ of the level $nl$

$$\varphi_{nl}(\vec{r})\big|_{r\to\infty} \approx \alpha_{nl}^{3/2}(\alpha_{nl}r)^{\frac{1}{\alpha_{nl}}-1}e^{-\alpha_{nl}r}, \tag{14}$$

where $\alpha_{nl} = \sqrt{2|E_{nl}|}$.

The asymptotic with account of electron exchange is determined, contrary to (14), not by $E_{nl}$, but by the smallest in absolute value binding energies, if states with higher principal quantum numbers than $n$ are occupied.

Let us consider the asymptotic of the one-particle HF wave function, taking for simplicity a two-level "atom", with one inner $i$ and the other outer $o$ electron and consider the equation for the inner state with wave function $\varphi_i(x)$ in order to see how it is modified by the exchange with the outer electron. Instead of (1) we have

$$\left[-\frac{\Delta}{2} - \frac{Z}{r} + \int \rho(x')\frac{1}{|\vec{r}'-\vec{r}|}dx'\right]\varphi_i(x) - \int \varphi_o^*(x')\frac{dx'}{|\vec{r}'-\vec{r}|}\varphi_i(x')\varphi_o(x) = -|E_i|\varphi_i(x). \tag{15}$$

If the last term in the left hand side of (15) is neglected, the asymptotic (14) is correct.

At large distances the exchange term $\Re(r)$ behaves as

$$\Re(r)\big|_{r\to\infty} = \int \varphi_o^*(x')\frac{dx'}{|\vec{r}'-\vec{r}|}\varphi_i(x')\varphi_o(x)\bigg|_{r\to\infty} \approx \frac{1}{r^2}\varphi_o(x)\int \varphi_o^*(x')(\vec{r}'\vec{n})\varphi_i(x')dx' \equiv \frac{C_{n_o}}{r^2}\varphi_o(x). \tag{16}$$

where $\vec{n}$ is the unit vector in the direction $\vec{r}$ and index $n$ stands for the principal quantum number of outer electron.

It is evident from (16) that located inside the "atom" wave function $\varphi_i(\vec{r})$ mixes with $\varphi_o(x)$ of much bigger radius. If to consider for concreteness $i = 1s$ then $\varphi_{1s}(\vec{r}) \approx \alpha_{1s}^{3/2}e^{-\alpha_{1s}r}$, where $\alpha_{1s} = \sqrt{2|E_{1s}|}$. As it follows from (16), s-state can be mixed with p-states only. Then using (15) and (16), we obtain the following expression for the asymptotic of $\varphi_i(\vec{r})$ [18]:

$$\varphi_i(\vec{r})\big|_{r\to\infty} \approx \alpha_i^{3/2}(\alpha_i r)^{\frac{1}{\alpha_i}-1}e^{-\alpha_i r} + \frac{C_{n_o}}{(\alpha_i r)^2}\alpha_{n_o l}^{3/2}(\alpha_{n_o l}r)^{\frac{1}{\alpha_{n_o l}}-1}e^{-\alpha_{nl}r}. \tag{17}$$

If $\alpha_i$ is considerably bigger than $\alpha_{n_o}$, i.e. if the energy levels are well separated, the first term in (17) can be neglected leading to

$$\varphi_i(\vec{r})\big|_{r\to\infty} \approx \frac{C_{n_o}}{(\alpha_i r)^2} \alpha_{n_o l}^{3/2} (\alpha_{n_o l} r)^{\frac{1}{\alpha_{n_o l}}-1} e^{-\alpha_{nl} r}, \tag{18}$$

thus completely modifying the asymptotic as compared to the case without exchange. Note that for pure hydrogen field $\alpha_{nl} = 1/n$ and $E_n = -1/2n^2$.

Thus, we demonstrated analytically that the asymptotic of any one-electron HF occupied state wave function is determined not by the state's binding energy $E_i$ but can be much bigger, $\sim \exp(-\sqrt{2|E_{\min}|}r)$, where $E_{\min}$ is the energy of the outermost particle. If there are several outer levels, the effect of exchange is determined by the following expression

$$\varphi_i(\vec{r})\big|_{r\to\infty} \approx \sum_{\text{All outer } n_o} \frac{C_{n_o}}{(\alpha_i r)^2} \alpha_{n_o l}^{3/2} (\alpha_{n_o l} r)^{\frac{1}{\alpha_{n_o l}}-1} e^{-\alpha_{n_o l} r}, \tag{19}$$

that for $N_o$ outer electrons enhances the exchange influence by factor $N_o$ in the amplitude. So, the role of exchange can be enhanced by exciting the outer electrons to states with smaller in absolute value energies.

Let us show why long-tail corrections to the inner one-electron wave functions due to exchange with outer electrons modify the probability of their elimination from an atom by a strong electric field, of which a concrete example can serve a high intensity (about $10^{18\text{-}20}$ and higher Watts/cm$^2$) and low frequency ($\omega \ll I$) laser beam [15].

The combination of static external and atomic field is depicted in Fig. 4. Let us for simplicity still treat a two-level atom. The probability to be ionized by the static field $\vec{E}$ for electrons $i$ and $o$ is determined by the probability to find corresponding electrons at points $r_i = I_i/E$ and $r_{n_o} = I_{n_o} \ll r_i$. This is given by square module of the corresponding wave functions at points $r_i$ and $r_{n_o}$. Assume that these points belong already to the asymptotic region for the wave functions. With account of inter-electron Coulomb interaction and exchange, the wave function of inner electron is given by (17) that lead to the following decay probability of the inner level "$i$"

$$"i" \quad |\varphi_{i,ex}(r_i)|^2 \approx \left| \alpha_i^{3/2} (\alpha_i r_i)^{\frac{1}{\alpha_i}-1} e^{-\alpha_{il} r} + \frac{C_{n_o}}{(\alpha_i r_i)^2} \alpha_{n_o l}^{3/2} (\alpha_{n_o l} r_i)^{\frac{1}{\alpha_{n_o l}}-1} e^{-\alpha_{n_o l} r_i} \right|^2. \tag{20}$$

For deep levels the contribution of the first term is negligible, so that the penetration of the electron out of the atom is given by expression

$$|\varphi_{i,ex}(r_i)|^2 \approx \frac{\alpha_{n_o}^3 C_{n_o}^2}{(\alpha_i r_i)^2} (\alpha_{n_o l} r_i)^{\frac{2}{\alpha_{n_o l}}-2} e^{-2\alpha_{n_o l} r_i}. \tag{21}$$

The enhancement factor $\eta$ due to inclusion of the Fock term into the one-electron wave function of the inner electron is determined by the ratio of (21) to the expression with neglect of the second term (20)

$$\eta \equiv \frac{|\phi_{i,ex}(r_i)|^2}{|\phi_i(r_i)|^2} = \frac{\alpha_{n_o l}^3 C_{n_o}^2}{(\alpha_i r_i)^2}(\alpha_{n_o} r_i)^{2\left(\frac{1}{\alpha_i} - \frac{1}{\alpha_{n_o l}}\right)} e^{2(\alpha_i - \alpha_{n_o l})r_i}. \qquad (22)$$

If there are $N_o$ outer electrons, the factor $\eta$ in accord with (12) acquire an additional enhancement factor $N_o^2$.

To illustrate the size of $\eta$, let us consider a numerical example, in which the inner electron binding energy $I_i$ is 5, while the outer electron binding energy $I_o$ is 1/2 and the external field intensity **E** is one[4]. Then the factor $\eta$ is of the order of $5.64 \times 10^{13}$, while for the same field and levels energy 1 and 10 atomic units, respectively, one has $7.86 \times 10^{38}$!

These tremendously big numbers are consequences of extremely small probability to eliminate an inner electron without exchange with the outer. Therefore it is more interesting and instructive to compare the ratio $\tau$ of inner to outer electron ionization probabilities when the exchange between outer and inner electrons is taken into account. This ratio is given by the following expression

$$\tau \equiv \frac{|\varphi_{i,ex}(r_i)|^2}{|\varphi_o(r_o)|^2} = \frac{\alpha_{n_o l}^3 C_{n_o}^2}{(\alpha_i r_i)^4}(\alpha_{n_o l} r_i / r_{n_o})^{\frac{2}{\alpha_{n_o l}} - 2} e^{2\alpha_{n_o l}(r_{n_o} - r_i)} \approx$$

$$\approx \frac{\alpha_{n_o l}^3 C_{n_o}^2}{(\alpha_i r_i)^4}(\alpha_{n_o l} r_i / r_{n_o})^{\frac{2}{\alpha_{n_o l}} - 2} e^{-2\sqrt{2I_{n_o l} I_i}/E} \sim E^4 \exp(-2\sqrt{2I_{n_o l} I_i}/E) \qquad (23)$$

For the considered examples of the energies of two levels, it is obtained for $\tau \approx 4.49 \times 10^{-5}$ and $5.01 \times 10^{-13}$, respectively. For the first case the ratio is not too small.

Qualitatively, it looks like the exchange admixture of outer electron literally "drags out" the inner electron off the ionized atom.

As it was mentioned before, if it is $N_o$ outer electrons, for which the coefficient $C_{n_o}$ is non-zero, this ratio is increased by additional factor $N_o^2$. It seems that this dependence was really observed in a number of investigations (see e.g. [11]) of multiple photoionization of noble-gas clusters by high intensity laser beam. In this studies a prominent amount of photons with energies of several hundreds of eV were found signaling the possibility that vacancies in inner shells were generated during laser-cluster interaction. The intensity of such processes in clusters could be a direct consequence of presence of very many outer electrons in clusters, contrary to the case of isolated atoms.

Note that exchange effects could be strongly enhanced even if the target atom exists in an exited state for a relatively short time. Therefore, presence of strong atomic resonances, e.g. Giant, at laser frequency can enhance the multiple ionization probability considerably. Perhaps this is the reason why in photoionization by free-electron laser an abundance of multiply charged ions, with degree of ionization up to twenty-one were found [10].

It was demonstrated recently both numerically and analytically in [16, 17] that the Fock exchange leads in fact to non-exponential instead of exponential barrier penetration probability. This is seen qualitatively already from (22): if $\alpha_{n_o l} r_o \sim 1$, the second term presents power decrease of the barrier penetration probability.

As appropriate objects for HF equations, however much more difficult for calculations, are atoms imbedded in condensed matter objects, clusters, fullerenes or endohedrals.

---
[4] That corresponds to the field intensity $10^{16}$ W/cm$^2$.

However, since they have much more outer electrons than an isolated atom, the wave function of an inner electron is modified stronger than in atoms.

One can expect traces of the discussed effects in atomic collisions in the strong laser field, while temporarily effectively exchanging objects are formed.

Giving the approximate nature of HF approach, it is essential to know whether account of electron correlations preserves or destroys the Fock's exchange contribution. It is possible to show using the example of infinite electron gas that non-locality is preserved but in case of high density electron gas noticeably diminished. In [17] arguments are presented, that correlations in atoms correct the inner electron asymptotic but by terms of higher powers in $1/r$ than the Fock term.

As I recently learned the approach developed in this section was criticized in [19]. I find this criticism not justified. The derivation of asymptotic (18) is transparent and directly follows from the non-simplified HF equation, while calculation in [19] are using a code that restricts the asymptotic to (14) and mistakenly uses as HF energies non consistently calculated data. This means unjustified inclusion into HF some terms that are definitely out of its frame. The statement made in [19] that the generally accepted approach in solving HF system of equations made some basis truncation is simply incorrect.

## 5. Conclusion and perspectives

We briefly outlined here some properties of fullerenes and endohedrals as big atoms with additional electron shells. We mainly qualitatively presented the picture of the response of this objects to photon beams. Main attention was concentrated on their resonance behavior that is most interesting for endohedrals that act as resonators for the photoelectrons eliminated from the inner atom and as an amplifier of the incoming electromagnetic radiation.

Such features of fullerenes and endohedrals as Giant, confinement and Giant endohedral resonances made them attractive and promising objects for studies of their interaction with high intensity lasers, both in the low and high frequency region.

A lot should be done in theoretical investigation of these objects in order to substitute very simple theoretical approaches employed here by more realistic ones that would account for the fullerenes and endohedrals structure more rigorously, but, desirable, not losing completely the achieved transparency in investigating the considered property. For instance, it is necessary to take into account that photoelectrons in endohedrals on their way out can ionize the fullerenes electrons thus increasing the total output of emitted electrons and photons.

It could be that the use of laser clarifies some other cooperative motions in the systems under consideration, such as e.g. excitation of oscillation of atom A in $A@C_{60}$.

Let me remind that $C_{60}$ is the simplest object of a fullerene, just as noble gases are the simplest "gest" to be stuffed inside from the theoretic point of view. It is the duty of theorist to more carefully investigate other fullerenes with or without other than noble gas atoms stuffed inside.

To summarize, I do feel that "big atoms" and their interaction with electromagnetic radiation is a very promising domain of research.

## Acknowledgements

The author is grateful for the financial assistance via the Israeli-Russian grant RFBR-MSTI 11-02-92484

**Figures**

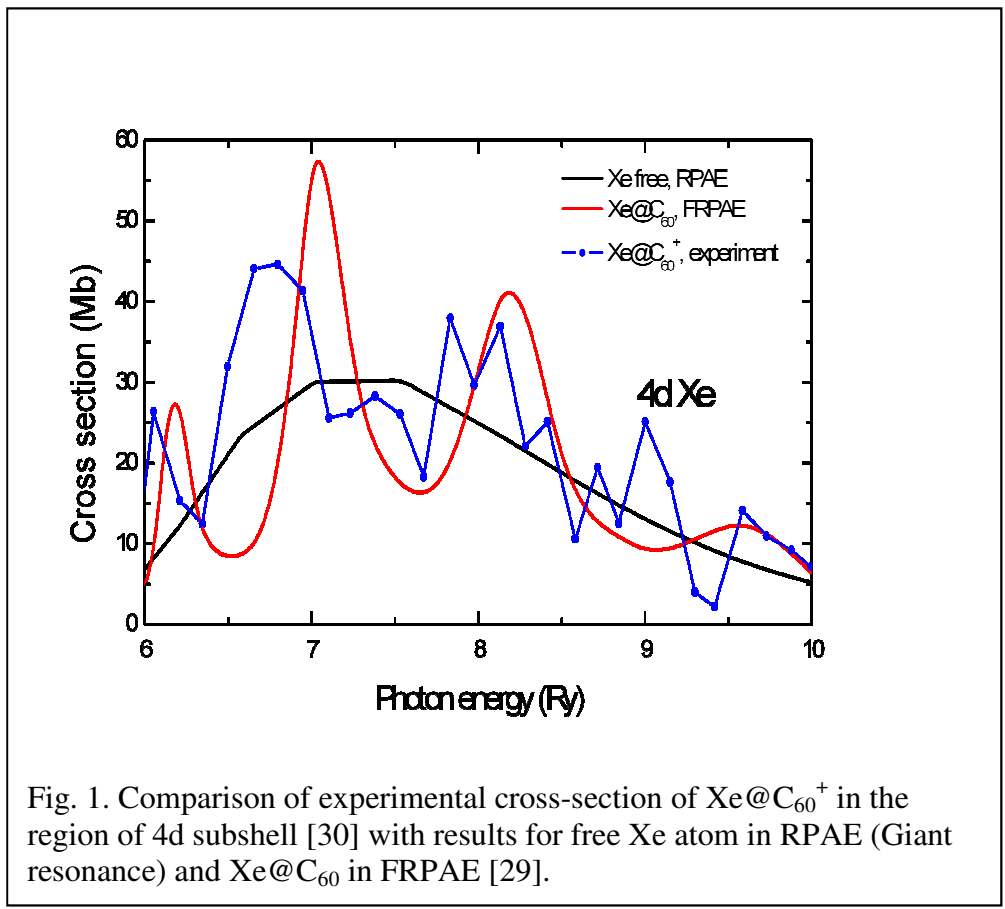

Fig. 1. Comparison of experimental cross-section of Xe@$C_{60}^+$ in the region of 4d subshell [30] with results for free Xe atom in RPAE (Giant resonance) and Xe@$C_{60}$ in FRPAE [29].

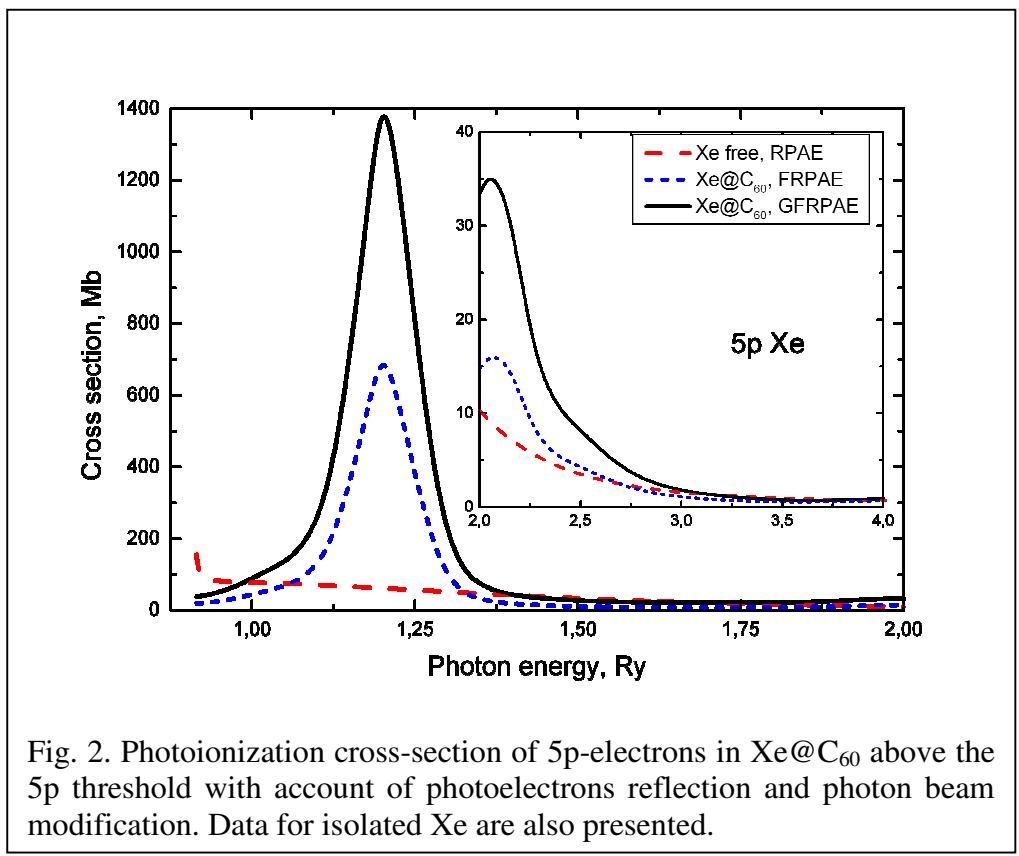

Fig. 2. Photoionization cross-section of 5p-electrons in Xe@$C_{60}$ above the 5p threshold with account of photoelectrons reflection and photon beam modification. Data for isolated Xe are also presented.

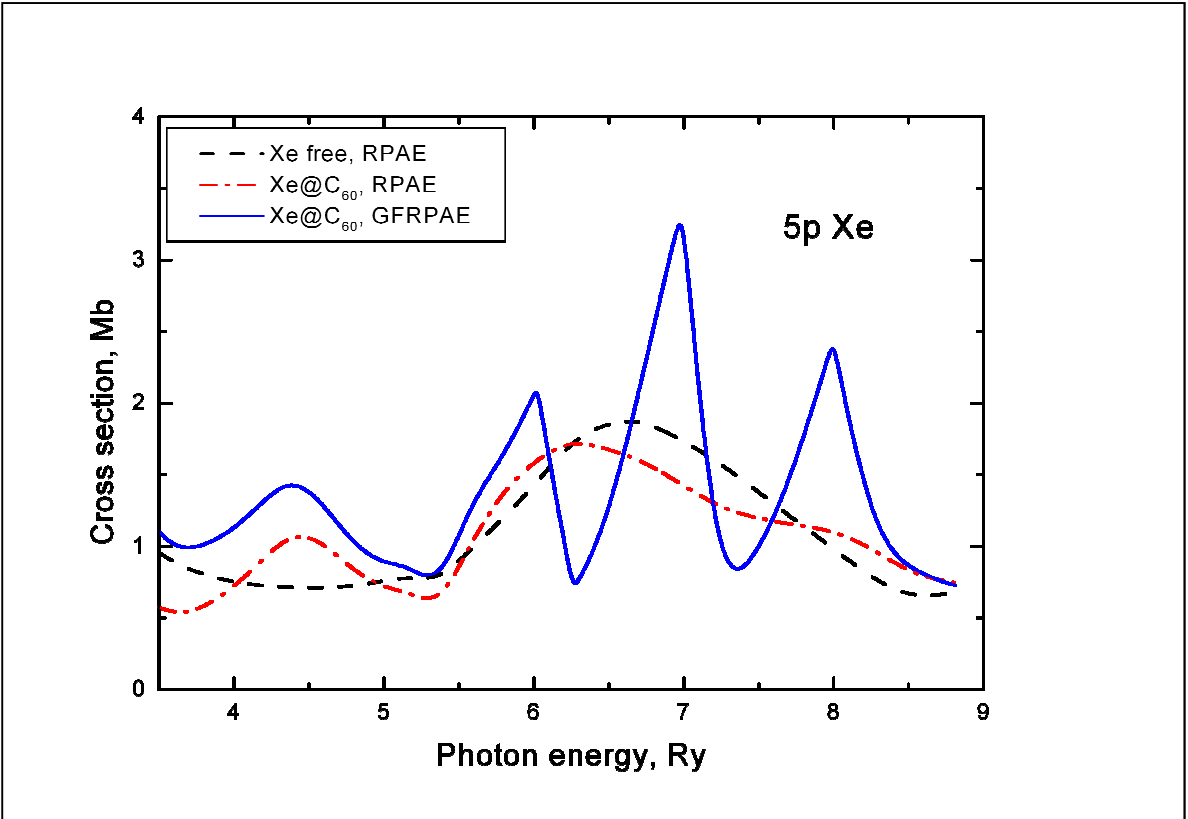

Fig. 3. Photoionization cross-section of 5p-electrons in Xe@C$_{60}$ near and above the 4d threshold with account of photoelectrons reflection and photon beam modification Data for isolated Xe are also presented.

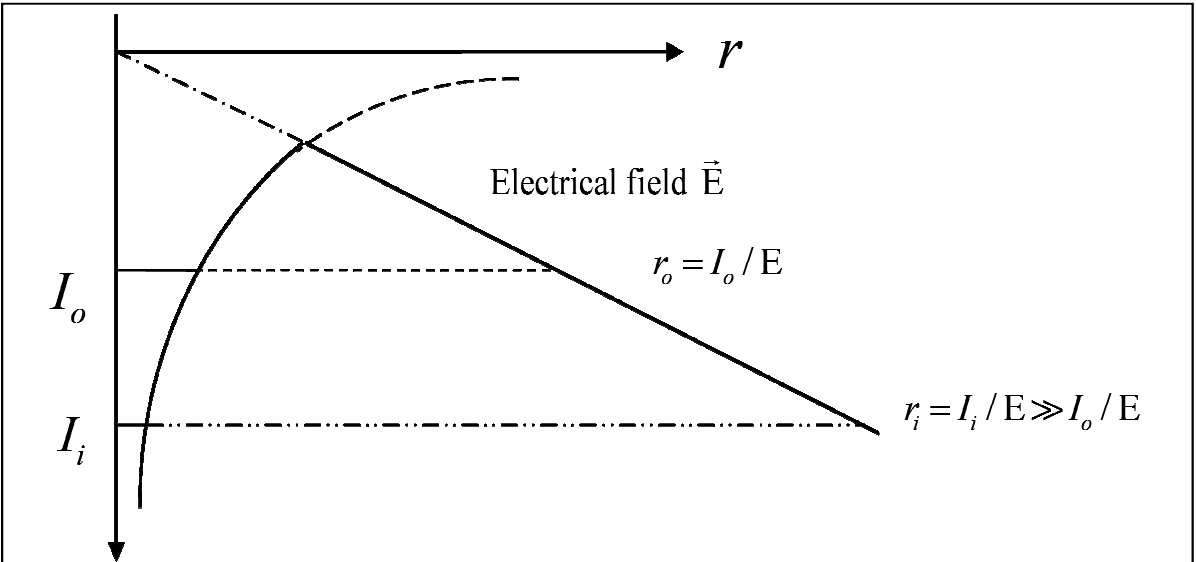

Fig. 4. Schematic representation of atomic and external electric field combination. Barriers for two, inner $i$ and outer $o$ atomic levels are demonstrated.